\documentclass[aps,preprint,
superscriptaddress,showpacs]{revtex4}
\usepackage{epsf}
\usepackage{amsmath}
\usepackage{multirow}
\begin{document}

\title{Freezing and Collapse of Flexible Polymers on Regular Lattices in Three Dimensions}
\author{Thomas Vogel}
\email[E-mail: ]{Thomas.Vogel@itp.uni-leipzig.de}
\affiliation{Institut f\"ur Theoretische Physik,
Universit\"at Leipzig, Postfach 100\,920, D-04009 Leipzig,\\
and Centre for Theoretical Sciences (NTZ), Emil-Fuchs-Stra{\ss}e 1, D-04105 Leipzig, Germany}
\author{Michael Bachmann}
\email[E-mail: ]{Michael.Bachmann@itp.uni-leipzig.de}
\affiliation{Institut f\"ur Theoretische Physik,
Universit\"at Leipzig, Postfach 100\,920, D-04009 Leipzig,\\
and Centre for Theoretical Sciences (NTZ), Emil-Fuchs-Stra{\ss}e 1, D-04105 Leipzig, Germany}
\affiliation{Computational Biology \& Biological Physics Group, 
Department of Theoretical Physics, Lund University,
S\"olvegatan 14A, SE-223\,62 Lund, Sweden}
\author{Wolfhard Janke}
\email[E-mail: ]{Wolfhard.Janke@itp.uni-leipzig.de}
\homepage[\\ Homepage: ]{http://www.physik.uni-leipzig.de/CQT.html}
\affiliation{Institut f\"ur Theoretische Physik,
Universit\"at Leipzig, Postfach 100\,920, D-04009 Leipzig,\\
and Centre for Theoretical Sciences (NTZ), Emil-Fuchs-Stra{\ss}e 1, D-04105 Leipzig, Germany}
\begin{abstract}
We analyze the crystallization and collapse transition of a simple model for
flexible polymer chains on simple cubic and face-centered cubic lattices by 
means of sophisticated chain-growth methods. In contrast to bond-fluctuation
polymer models in certain parameter ranges, where these two conformational
transitions were found to merge in the thermodynamic limit, we conclude from our 
results that the two transitions remain well-separated
in the limit of infinite chain lengths. The reason for this qualitatively 
distinct behavior is presumably due to the ultrashort 
attractive interaction range in the lattice models considered here.
\end{abstract}
\pacs{05.10.-a, 87.15.Aa, 87.15.Cc}
\maketitle
\section{\label{sec:intro}Introduction}
The analysis of conformational transitions a single polymer in solvent can 
experience is surprisingly difficult. In good solvent (or high temperatures), 
solvent molecules occupy binding sites of the polymer and, therefore, the 
probability of noncovalent bonds between attractive segments of the polymer 
is small. The dominating structures in this phase are dissolved or random coils.
Approaching the critical point at the $\Theta$ temperature, the polymer collapses 
and in a cooperative arrangement of the monomers, globular conformations are 
favorably formed. At the $\Theta$ point, which has already been studied over many
decades, the infinitely long polymer behaves like a Gaussian
chain, i.e., the effective repulsion due to the volume exclusion constraint 
is exactly balanced by the attractive monomer-monomer interaction. Below
the $\Theta$ temperature, the polymer enters the globular phase, where the influence of 
the solvent is small. Globules are very compact conformations, but there is
little internal structure, i.e., the globular phase is still entropy-dominated.
For this reason, a further transition towards low-degenerate energetic states is
expected to happen: the freezing or crystallization of the polymer. Since this transition
can be considered as a liquid-solid phase separation process, it is expected to
be of first order, in contrast to the $\Theta$ transition, which exhibits 
characteristics of a second-order phase transition~\cite{lifshitz1,khokhlov1}.

The complexity of this problem appears in the quantitative description of these
processes. From the analysis of the corresponding field theory~\cite{deGennes1} 
it is known that for the $\Theta$ transition
the upper critical dimension is $d_{\rm c}=3$, i.e., multiplicative and additive
logarithmic corrections to the
Gaussian scaling are expected and, indeed, predicted by field 
theory~\cite{duplantier1,duplantier2,stephen1,hager1}. However, until now neither experiments nor computer
simulations could convincingly provide evidence for these logarithmic corrections.
This not only regards analyses of different single-polymer 
models~\cite{grass1,whitt1,whitt2,grass2,binder1,binder1b,parsons1}, but also the related problem of 
critical mixing and unmixing in polymer solutions~\cite{binder2,grass3,pana1,yan1,anisimov1}.

In a remarkable recent study of a  bond-fluctuation polymer model, it was shown that,
depending on the intramolecular interaction range, collapse and freezing transition 
can fall together in the thermodynamic limit~\cite{binder1,binder1b}. This surprising
phenomenon is, however, not general. For an off-lattice bead-spring polymer with FENE 
(finitely extensible nonlinear elastic) bond potential and intra-monomer Lennard--Jones
interaction, for example, it could be shown that both transitions remain well separated
in the limit of infinitely long chains~\cite{parsons1}. 

In our study, we investigate collapse and freezing of a single homopolymer restricted
to simple cubic (sc) and face-centered cubic (fcc) lattices. We primarily focus on
the freezing transition, where comparatively little is known as most of the analytical
and computational studies
in the past were devoted to the controversially discussed collapse transition; see, e.g., 
Refs.~\cite{grass1,grass2,grass3,pana1,yan1,mccrack1,bruns1,kremer1,mei1,kremer2,taylor1,prell3}. 
A precise 
statistical analysis of the conformational space relevant in this low-temperature 
transition regime
is difficult as it is widely dominated by highly compact low-energy conformations which 
are entropically suppressed. Most promising for these studies appear sophisticated 
chain-growth methods based on Rosenbluth sampling~\cite{rosen1} combined with improved 
pruning-enrichment strategies~\cite{grass2,hsu1} which, in their original formulation, 
are particularly useful for the
sampling in the $\Theta$ regime. For the analysis of the freezing transition, we apply 
in our simulations more generalized contact-density variants~\cite{bj1,prell1}, which have 
proven to be very successful in the low-energy sampling of protein-like 
heteropolymers~\cite{bj1} and the adsorption of polymers and peptides to solid 
substrates~\cite{bj2,prell2}. The precision of these algorithms when applied to 
lattice polymers as in the present study, is manifested by unraveling even 
finite-length effects induced by symmetries of the underlying lattice. 

The rest of the paper is organized as follows. 
In Sect.~\ref{sec:models}, the lattice model for flexible polymers and 
the employed chain-growth methods, which allow for a precise statistical sampling even in the low-temperature
regime, are described. The conformational transitions the polymers experience on sc and fcc lattices 
are discussed in Sect.~\ref{sec:results}. Here, we also present our results for the scaling 
of the collapse transition temperature in comparison with various approaches known from the literature. 
Eventually, in Sect.~\ref{sec:summary}, the paper is concluded by a summary of our findings. 
\section{\label{sec:models}Model and methods}
For our studies, we employ the interacting self-avoiding walk (ISAW) model for lattice polymers.
In this model, the polymer chain is not allowed to cross itself, i.e., a lattice site can only be
occupied by a single monomer. In order to mimic the ``poor solvent'' behavior in the energetic 
regime, i.e., at low temperatures, nearest-neighbor contacts of nonadjacent monomers reduce 
the energy. Thus, the most compact conformations possess the lowest energy. Formally, the 
total energy of a conformation $\textbf{X}=\{\textbf{x}_1,\textbf{x}_2,\ldots, \textbf{x}_N \}$
of a chain with $N$ beads is simply given as
\begin{equation}
\label{eq:mod}
E(\textbf{X})=-\varepsilon_0 n_{\rm NN}(\textbf{X}),
\end{equation}
where $\varepsilon_0$ is an unimportant energy scale (which is set $\varepsilon_0\equiv 1$ in the 
following) and $n_{\rm NN}(\textbf{X})$ is the number of nearest-neighbor contacts between nonbonded
monomers. 

The total number of self-avoiding lattice conformations with $m=N-1$ bonds scales as~\cite{gutt1,gutt2}
\begin{equation}
C_m\sim \mu^mm^{\gamma-1},
\end{equation}
where $\mu$ is the effective coordination number of the lattice and $\gamma\approx 1.16$ a universal exponent.
For the sc lattice, the connectivity constant is $\mu_{\rm sc}\approx 4.684$ and in the fcc case
$\mu_{\rm fcc}\approx 10.036$~\cite{mac1,cara1,guida1,gutt3,chen1,vogel1,sbj1}. Due to this exponential growth of 
the number of conformations, the investigation of all conformational transitions a homopolymer of a given length 
can experience requires employing numerical methods being capable of estimating the density of states for
all possible energies with high precision. There exist mainly two strategies for generating and updating 
conformations within the stochastic search schemes typically used. Applying standard Markov chain Monte Carlo
methods, conformational updates include semilocal changes of bond orientations as, among others, 
corner and end flips, crankshaft moves, and more non-local updates such as pivot rotations. In our work,
we have used the alternative concept of chain growth. Depending on the lattice constraints, a new monomer
is tried to be attached to an end of the already existing chain until the total length or a ``dead end''
is reached, i.e., all neighbors are already occupied and, thus, the chain end is trapped. In an early 
approach,
the Rosenbluth method~\cite{rosen1}, firstly the number of free nearest neighbors $k$ for the 
possible placements of 
the $l$th monomer is determined.
Then, one of the possibilities (if any) is selected randomly. The peculiarity is that this algorithm introduces a
bias as the (athermal) probability of generating a certain chain conformation $\textbf{X}$ of length $N$, 
$p(\textbf{X})=\prod_{l=2}^{N} k^{-1}_l$, depends on the growth direction. Thus, identical walks
can possess different construction probabilities, if they, e.g., were grown ``forward'' or ``backward''. 
For correct statistics, this bias must be corrected by introducing Rosenbluth weights 
$w(\textbf{X})=p^{-1}(\textbf{X})$. 
Actually, this bias can be utilized to increase the efficiency of the method in generating self-avoiding
walks which is particularly useful for the 
ISAW model of lattice polymers. For this purpose it is convenient to introduce for each chain a thermal 
Rosenbluth weight 
$W(\textbf{X})=\prod_{l=2}^{N} k_l\exp\{-[E(\textbf{x}_{l})-E(\textbf{x}_{l-1})]/k_B T\}$ 
(where $k_B$ is the Boltzmann constant which we set to $k_B\equiv 1$ in our analysis). 
Since
this method is a clever kind of simple sampling, the partition sum can be estimated \textit{absolutely}
as $Z_N\approx\sum_{i=1}^{I_N} W(\textbf{X}_i)/M$, where $I_N$ is the number of successfully generated 
polymer chains of length $N$ in $M$ growth starts~\cite{grass2}. In principle, since 
$Z_N=\sum_{E}g_N(E)\exp(-E/k_BT)$, an absolute estimate of the density of 
states $g_N(E)$, i.e., the degeneracy 
of energetic states $E$, is then also known. This is particularly important for heteropolymers, where the 
ground-state degeneracy is considered as a measure for the stability of native folds of lattice 
proteins~\cite{bj1}. An essential improvement of the efficiency of this chain-growth approach was reached 
by the introduction of the pruned-enriched Rosenbluth method (PERM)~\cite{grass2} and its enhanced 
variants~\cite{hsu1}, which combine 
Rosenbluth chain growth with a 'Go with the Winners'~\cite{aldous1} strategy. 
In these algorithms, at each stage
of the growth process copies of the already existing chain segment are created and 
continue growing independently, 
if the accumulated Rosenbluth weight is larger than an upper threshold. 
If the weight falls below a lower bound, the
chain is pruned with a certain probability (which is typically 1/2). Otherwise, the growth of the single chain
simply continues. This method has frequently been applied in studies of the $\Theta$ 
point~\cite{grass1,grass2,prell3,grass3}. In our study, we use the nPERMss (new PERM with simple sampling)
variant~\cite{hsu1} for the simulation of $\Theta$ polymers with chain lengths of up to 32\,000 (sc)
and 4\,000 (fcc) monomers, respectively.

For the analysis of the conformational behavior below the $\Theta$ point, we use even more sophisticated,
generalized-ensemble variants which are independent of the temperature and yield an improved estimate 
for the density of states $g(E)$ within a single simulation. These algorithms combine 
PERM-based chain growth with 
multicanonical~\cite{bj1} or flat-histogram techniques~\cite{prell1}
and increase, in particular, the sampling of entropically suppressed (``rare'') conformations, which are,
for example, essential for the study of the freezing transition. Due to the much higher demands in this regime,
maximum chain lengths, for which precise results were reliably obtained, are $N=125$ (sc) and 56 (fcc),
respectively. 

For our statistical analysis, it is convenient to define energetic statistical expectation values via
the density of states, i.e., $\langle O(E)\rangle=\sum_{E}g_N(E)O(E)\exp(-E/k_BT)/Z_N$. 
The main results of our analysis are based on the peak structure of the specific heat which is defined
as $C_V(T)=d\langle E\rangle/dT=\left(\langle E^2\rangle-\langle E\rangle^2\right)/k_BT^2$. 
\section{\label{sec:results}Results and discussion}
It was recently found for a bond-fluctuation model with inter-monomeric interaction radius 
$r=\sqrt{6}$ that in the infinite chain-length limit collapse and freezing are 
indistinguishable phase transitions appearing at the same
temperature (the $\Theta$ temperature $T_\Theta$)~\cite{binder1,binder1b}. 
In a bead-spring FENE model analysis~\cite{parsons1},
this phenomenon could not be observed: Both transitions exist in the thermodynamic limit
and the crossover peaks in the specific heat remain well-separated. The same observation
was made independently in Ref.~\cite{binder1b} for the bond-fluctuation model
with increased interaction range.
In the following, we perform 
for the lattice polymer model (\ref{eq:mod}) a detailed analysis of these transitions on regular sc and fcc 
lattices and discuss the expected behavior in the thermodynamic limit.
\subsection{\label{sec:peak}The expected peak structure of the specific heat}
\begin{figure}
\centerline{\epsfxsize=8.8cm \epsfbox{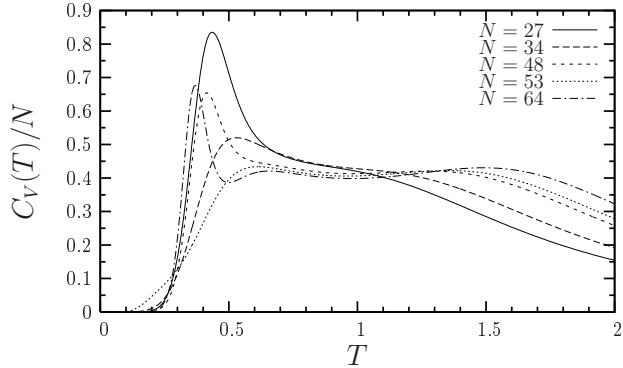}}
\caption{\label{fig:1} Examples of specific-heat curves (per monomer) for a few exemplified 
short homopolymers on the
sc lattice. Absolute errors (not shown) are smaller than $0.03$ in the vicinity of the
low-temperature peaks and smaller than $10^{-5}$ in the onset of the $\Theta$-transition region 
near $T\approx1.5$.}
\end{figure}
\begin{figure}
\centerline{\epsfxsize = 8.8cm \epsfbox{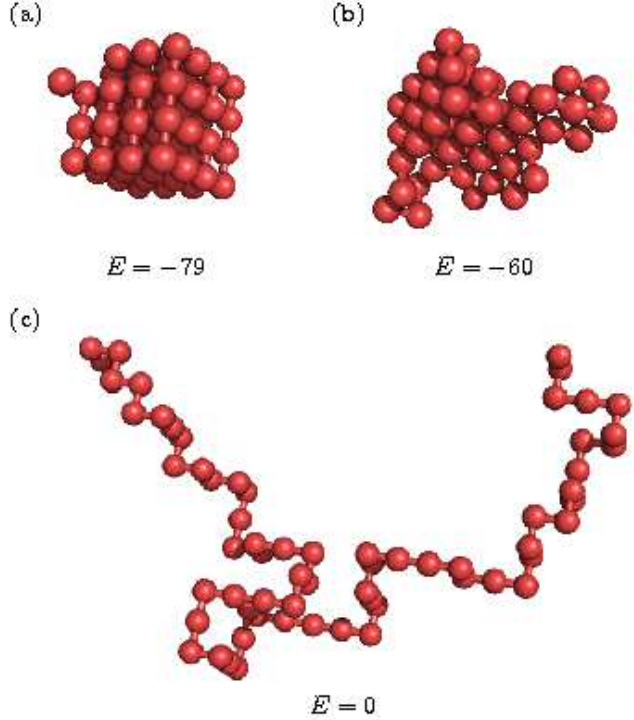}}
\caption{\label{fig:8}
Representative conformations of a 64-mer in the different pseudophases: (a) Excitation
from the perfect $4\times 4\times 4$ cubic ground state (not shown, $E=-81$) to
the first excited crystal state, (b) transition towards globular states, and 
(c) dissolution into random-coil conformations.}
\end{figure}
Statistical fluctuations of the energy, as expressed by the specific
heat, can signalize thermodynamic activity. Peaks of the specific heat as a function of temperature
are indicators for transitions or crossovers between physically different macrostates of the
system. In the thermodynamic limit, the collective activity, which influences typically most
of the system particles, corresponds to thermodynamic phase transitions. For a flexible polymer, three 
main phases are expected: The random-coil phase for temperatures $T>T_\Theta$, where conformations are 
unstructured and dissolved; the globular phase in the temperature interval $T_m<T<T_\Theta$
($T_m$: melting temperature) with condensed, but unstructured (``liquid'') conformations dominating; 
and for $T<T_m$ the ``solid'' phase characterized by locally crystalline or amorphous 
metastable structures. In computer simulations, only polymers of finite length are accessible and,
therefore, the specific heat possesses typically a less pronounced peak structure, as finite-length 
effects can induce additional signals of structural activity and shift the transition temperatures.
These effects, which are typically connected with surface-reducing monomer rearrangements, 
are even amplified by steric constraints in lattice models as used in our study. 
Although these pseudotransitions are undesired in the analysis of the thermodynamic transitions,
their importance in realistic systems is currently increasing with the high-resolution equipment 
available in experiment
and technology. The miniaturization of electronic circuits on polymer basis and possible nanosensory 
applications in biomedicine will, therefore, require a more emphasized analysis of the finite-length 
effects in the future.
\subsection{Simple-cubic lattice polymers}
Figure~\ref{fig:1} shows typical examples of specific heats for very short chains on the sc lattice
and documents the difficulty of identifying the phase structure of flexible homopolymers. The 27-mer
exhibits only a single dominating peak -- which is actually only an sc lattice effect. The reason is
that the ground states are cubic ($3\times 3\times 3$) and the energy gap towards the first excited 
states is $\Delta E=2$~\cite{rem1}. Actually, also the most pronounced peaks for 
$N=48$ ($4\times 4\times 3$) and
$N=64$ ($4\times 4\times 4$) are due to the excitation of perfectly cuboid and cubic ground states,
respectively. The first
significant onset of the collapse transition is seen for the 48-mer close to $T\approx 1.4$.
A clear discrimination between the excitation and the melting transition is virtually impossible.
In these examples, solely for $N=64$ three separate peaks are present. 
The plots in Figs.~\ref{fig:8}(a)--\ref{fig:8}(c) show representative conformations 
in the different pseudophases
of the 64-mer. Due to the energy gap, the excitations of the cubic ground state 
with energy $E=-81$ (not shown) to conformations with $E=-79$ [Fig.~\ref{fig:8}(a)] result in 
a pseudotransition which is represented by the first specific-heat peak in Fig.~\ref{fig:1}. 
The second less-pronounced peak in Fig.~\ref{fig:1} around $T\approx 0.6-0.7$ signalizes 
the melting into globular structures, whereas
at still higher temperatures $T\approx 1.5$ the well-known collapse peak indicates the dissolution into
the random-coil phase.

\begin{figure}
\centerline{\epsfxsize=8.8cm \epsfbox{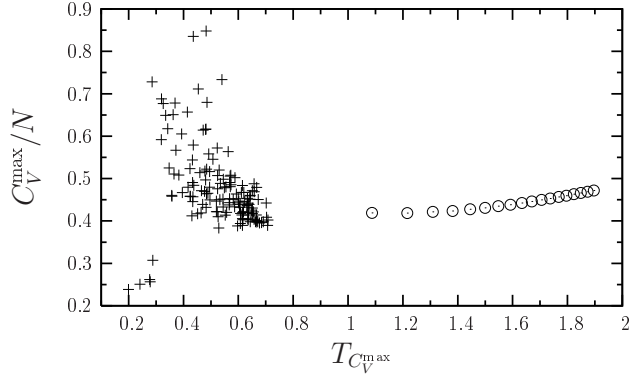}}
\caption{\label{fig:2}Map of specific-heat maxima for several chain lengths taken from the 
interval $N\in [8,125]$. 
Circles ($\odot$) symbolize the peaks (if any) identified as signals of the collapse ($T_{C_V^{\rm max}}>1$). 
The low-temperature peaks ($+$) belong to the excitation/freezing transitions ($T_{C_V^{\rm max}}<0.8$).
The group of points in the lower left corner corresponds to polymers with $N_c+1$ monomers, 
where $N_c$ denotes the ``magic'' lengths allowing for cubic or cuboid ground-state conformations
(see Fig.~\ref{fig:3} and text).}
\end{figure}
The distribution of the maximum values of the specific heat $C_V^{\rm max}$ with respect to the 
maximum temperatures $T_{C_V^{\rm max}}$ is shown in Fig.~\ref{fig:2}. Not surprisingly, the peaks 
belonging to the excitation and freezing transitions ($+$) appear to be 
irregularly ``scattered'' in the
low-temperature interval $0<T_{C_V^{\rm max}}<0.8$. The height of the peaks indicating the 
collapse transition of the finite-length polymers ($\odot$) is, on the other hand, monotonously 
increasing with the collapse-peak temperature. 
\begin{figure}
\centerline{\epsfxsize = 8.8cm \epsfbox{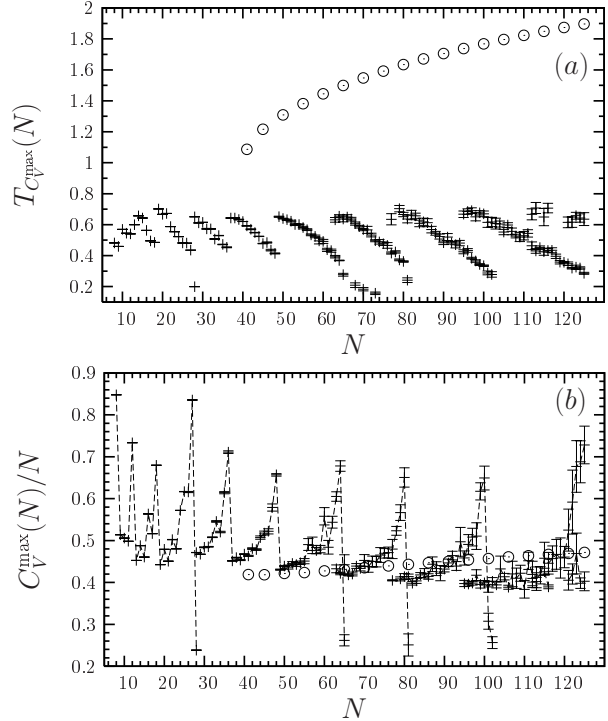}}
\caption{\label{fig:3}
(a) Collapse ($\odot$) and crystallization/excitation ($+$) peak temperatures 
of the specific heat for all chain lengths in the interval $N\in [8,125]$, 
(b) values of the specific-heat maxima in the same interval. Error bars for the collapse
transition data (not shown) are much smaller than the symbol size. $\Theta$ peaks appear starting
from $N=41$. For the sake of clarity, not all intermediate $\Theta$ data points are shown
(only for $N=41,45,50,\ldots$).}
\end{figure}
\begin{figure*}
\centerline{\epsfxsize = 17.6cm \epsfbox{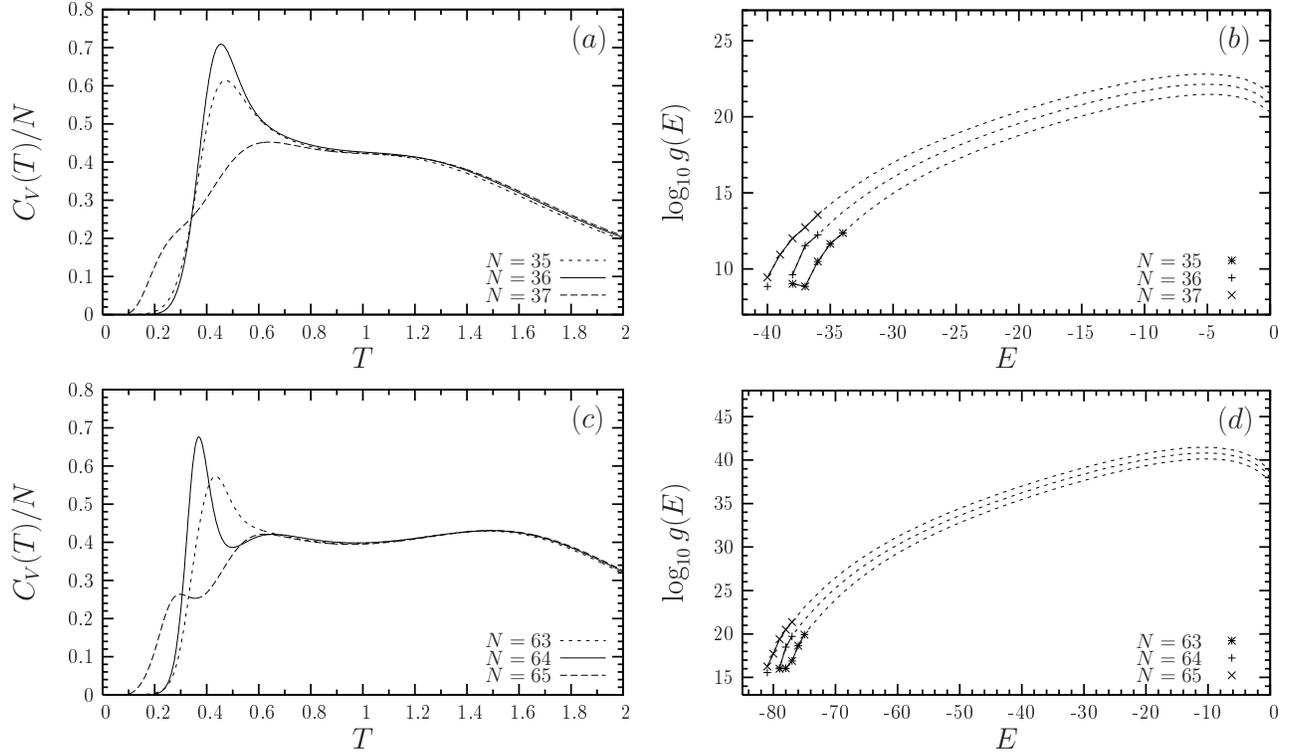}}
\caption{\label{fig:4}
Examples of specific heats for polymers with (a) $N_c=4\times 3\times 3=36$ and $N_c\pm 1$ 
and (c) $N_c=4\times 4\times 4=64$ and $N_c\pm 1$ monomers. In (b) and (d), the respective
densities of states are shown (lines are only guides to the eye). Symbols 
$*$, $+$ and $\times$ 
emphasize the lowest-energy states. Note the energy gaps between the ground and the first
excited states in the compact cases $N_c=36,64$ and the dip for $N=35,63$.}
\end{figure*}

Figures~\ref{fig:3}(a) and~\ref{fig:3}(b), showing the respective chain-length dependence 
of the maximum temperatures and maximum specific-heat values, reveal a more systematic
picture. At least from the results for the short chains shown, general scaling properties
for the freezing transition cannot be read off at all. The reason is that the low-temperature
behavior of these short chains is widely governed by lattice effects. This is clearly seen 
by the ``sawtooth'' segments. Whenever the sc chain possesses a ``magic'' length $N_c$ such that the
ground state is cubic or cuboid 
(i.e., $N_c\in{\cal N}_c=\{8,12,18,27,36,48,64,80,100,125,\dots\}$),
the energy gap $\Delta E=2$ between the ground-state conformation and the first excited state
entails a virtual energetic barrier which results in an excitation transition. 
Since entropy is less
relevant in this regime, this energetic effect is not ``averaged out'' and, therefore, causes
a pronounced peak in the specific heat [see Fig.~\ref{fig:3}(b)] at comparatively low 
temperatures [Fig.~\ref{fig:3}(a)]. This peculiar sc lattice effect vanishes widely by
increasing the length by unity, i.e., for chain lengths $N_c+1$. In this case, the excitation
peak either vanishes or remains as a remnant of less thermodynamic significance. The latter
appears particularly in those cases, where $N=N_c+1$ with $N_c=L^3$ (with $L$ being any 
positive integer) is a 
chain length allowing for perfectly cubic ground states. 
Increasing the polymer length further, the freezing peak dominates at low temperatures. Its peak 
increases with the chain length, whereas the peak temperature decreases. Actually, with increasing
chain length, the character of the transition converts from freezing to excitation, i.e., the entropic freedom that
still accompanies the melting/freezing process decreases with increasing chain length. 
In other words, cooperativity is lost: only a small fraction of monomers -- residing in the 
surface hull -- is entropically sufficiently flexible to compete the energetic gain of
highly compact conformations. This flexibility is reduced the more, the closer the chain
length $N$ approaches a number in the ``magic'' set ${\cal N}_c$.
If the next length belonging to ${\cal N}_c$ is reached, the next discontinuity in 
the monotonic behavior occurs. Since noticeable ``jumps'' are only present for chain lengths
whose ground states are close to cubes ($N_c=L^3$) or cuboids with $N_c=L^2(L\pm 1)$, 
the length of the branches in between scales with $\Delta N_c\sim L^2\sim N_c^{2/3}$. Therefore, only for very long
chains on the sc lattice, for which, however, a precise analysis of the low-temperature behavior is 
extremely difficult, a reasonable scaling analysis for $T_{C_V^{\rm max}}(N)$ 
and $C_V^{\rm max}(N)$ could be performed. 

\begin{figure}
\centerline{\epsfxsize = 8.8cm \epsfbox{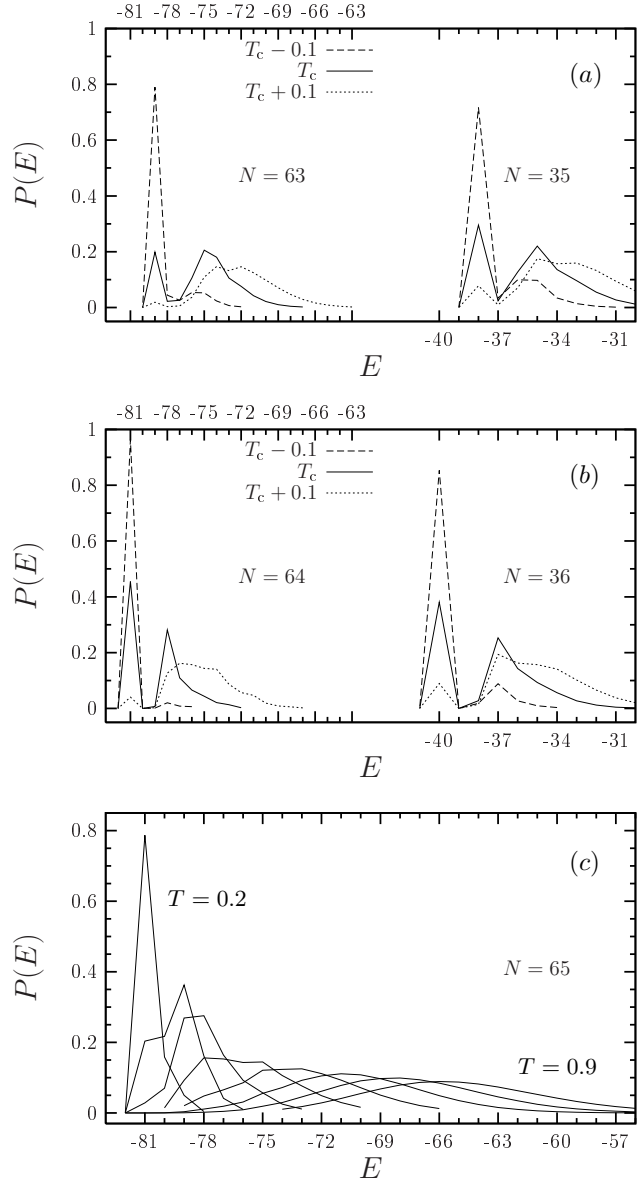}}
\caption{\label{fig:5}
Energy distributions $P(E)$ at low temperatures for sc lattice polymers with (a) $N=N_c-1=35$ and
$63$, (b) $N=N_c=36$ and $64$ monomers. In (c), $P(E)$ is shown for the chain with length $N=N_c+1=65$
for several temperatures $T=0.2,0.3,\ldots,0.9$. The distributions for 
$N=37$ (not shown) are similar. Note that lines are only guides to the eyes.}
\end{figure}

Exemplified for chains of lengths $N=N_c-1,N_c,$ and $N_c+1$ with $N_c=36,64$, specific heats and 
densities of states are shown in Figs.~\ref{fig:4}(a)--\ref{fig:4}(d), which exhibit the length-dependent
characteristic properties discussed above. While in Fig.~\ref{fig:4}(a) for chain lengths 
around $N=36$ only some low-temperature activity is visible and the collapse transition is vaguely
indicated by a broad shoulder, the transition characteristics are better resolved for $N=64$
shown in Fig.~\ref{fig:4}(c).
The most pronounced low-temperature peak of the
specific heat of the 64-mer, 
for example, is the excitation peak, the second peak belongs to the
freezing transition, and the third, still very shallow peak signals the collapse 
transition. The low-temperature behaviors
of the 63-mer and the 65-mer are quite different: While the low-temperature peak of the 
63-mer close to $T\approx 0.4$ is due to excitation as is indicated by the $E=-79$ ``dip'' in the
density of states in Fig.~\ref{fig:4}(d) [similar to the 35-mer in Fig.~\ref{fig:4}(b) at
$E=-38$], the relevant peak for the 65-mer is the
freezing peak close to $T\approx 0.6$. In this case, the excitation is of much less
relevance (although it is still reflected by a small peak near $T\approx 0.3$). This is a 
consequence of the missing convex lowest-energy dip in the density of states (or microcanonical
entropy). The convex monotony is a signal of a strong first-order phase separation~\cite{gross1,jbj1}. 
This is confirmed by analyzing the canonical energy distributions for
the examples $N=N_c-1,N_c,$ and $N_c+1$ with $N_c=36,64$ shown in Figs.~\ref{fig:5}(a)--\ref{fig:5}(c)
for temperatures close to the respective excitation and freezing transitions.
For $N_c-1=35,63$ [Fig.~\ref{fig:5}(a)], the pronounced excitation transition is
expressed by the respective double peaks with the strong gap in between, which
are for the polymers with chain lengths $N_c=36,64$ [Fig.~\ref{fig:5}(b)] due to
the energy gap between ground state and first excited state. This induces the 
first-order-like character of this pseudotransition. The energy distributions 
for various temperatures shown in Fig.~\ref{fig:5}(c) for the case $N=N_c+1=65$
do not exhibit, on the other hand, pronounced double-peaked shapes. The excitation 
transition at extremely low energies is still weakly present as a small 
shoulder in the distribution at the corresponding temperature. The freezing transition 
is associated with slightly larger energies (and temperature) and visible in the 
distribution with a weak tendency to a double-peaked shape.

The collapse transitions of the finite-length polymers are not affected by the intricate
low-energy conformations on the sc lattice and exhibit a continuous monotony. This
will be analyzed in Sect.~\ref{sec:theta} in more detail. 
\subsection{\label{sec:fcc}Polymers on the fcc lattice}
\begin{figure}
\centerline{\epsfxsize = 8.8cm \epsfbox{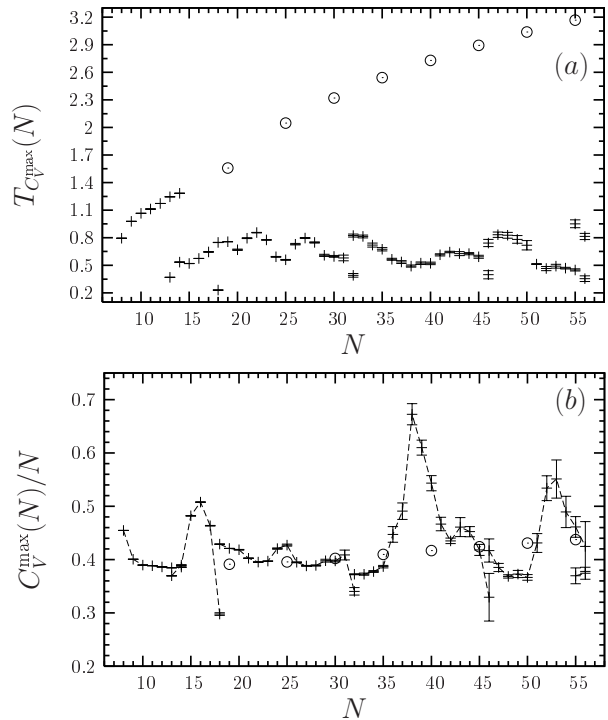}}
\caption{\label{fig:fcc1}
Peak temperatures (a) and peak values (b) of the specific heat for all chain lengths
$N=8,\ldots,56$ of polymers on the fcc lattice. Circles ($\odot$) symbolize the collapse peaks 
and low-temperature peaks ($+$) signalize the excitation/freezing transitions.
The error bars for the collapse transition are typically much smaller than the symbol size. 
Only for the freezing transition of 
longer chains, the statistical uncertainties are a little bit larger and visible in the plots.
$\Theta$ peaks appear starting
from $N=19$. For clarity, $\Theta$ data points are only shown for $N=19,25,30,\ldots$.
}
\end{figure}

The general behavior of polymers on the fcc lattice is comparable to what we found 
for the sc polymers. The main difference is that excitations play only a minor 
role, and the freezing transition dominates the conformational behavior 
of the fcc polymers at low temperatures. Nonetheless, finite-length effects are still
apparent as can be seen in the chain-length dependence of the peak temperatures and peak values
of the specific heats plotted in Fig.~\ref{fig:fcc1}(a) and Fig.~\ref{fig:fcc1}(b), respectively. 
Figure~\ref{fig:fcc1}(a) shows that the locations of the freezing and collapse transitions
clearly deviate with increasing chain lengths and we hence can conclude that also for fcc polymers
there is no obvious indication that freezing and collapse 
could fall together in the thermodynamic limit.

\begin{figure}
\centerline{\epsfxsize = 8.8cm \epsfbox{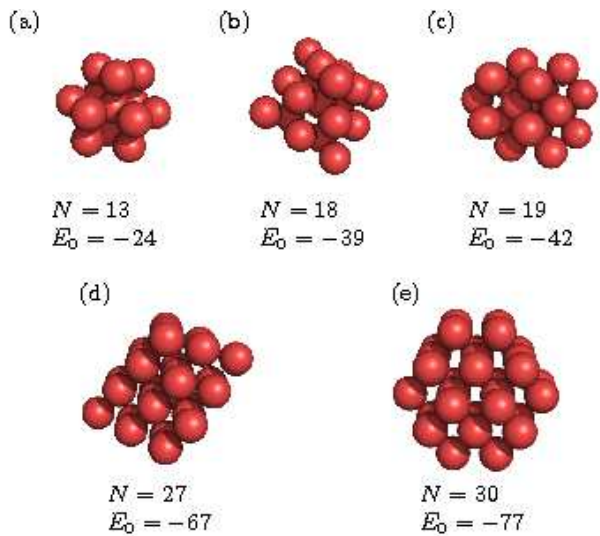}}
\caption{\label{fig:fcc2}
Ground-state conformations and energies of the (a) $13$-, (b) $18$-, (c) $19$-, 
(d) $27$-, and (e) $30$-mer on the fcc-lattice (bonds not shown).}
\end{figure}

Similar to the sc polymers, the finite-length effects at very low temperatures are 
apparently caused by the usual compromise 
between maximum compactness, i.e., maximum number of energetic (nearest-neighbor) contacts, 
and steric constraints of the underlying rigid
lattice. The effects are smaller than in the case of the sc lattice, as there are no 
obvious ``magic'' topologies in the fcc case. Ground-state conformations for a few
small polymers on the fcc lattice are shown in Fig.~\ref{fig:fcc2}. The general tendency is that
the lowest-energy conformations consist of layers of net planes with (111) orientation, i.e.,
the layers themselves possess triangular pattern with side lengths equal to the
fcc nearest-neighbor distance $\sqrt{2}$ (in units of the lattice constant). This is not surprising, 
as these conformations are tightly packed which ensures a maximum number of nearest-neighbor
contacts and, therefore, lowest conformational energy. An obvious example is the ground-state 
conformation of the 13-mer as shown in Fig.~\ref{fig:fcc2}(a) which corresponds to the intuitive
guess for the most closely packed structure on an fcc lattice: a monomer with its 12 nearest neighbors
(``3--7--3'' layer structure).
A simple contact counting yields 36 nearest-neighbor contacts which, by subtracting the $N-1=12$ 
covalent (nonenergetic) bonds, is equivalent to an energy $E=-24$. 
However, this lowest-energy conformation is degenerate. There is another conformation (not shown)
consisting of only two ``layers'', one containing 6 (a triangle) 
and the other 7 (a hexagon) monomers (``6--7'' structure), with the same 
number of contacts. 

A special case is the 18-mer. As Fig.~\ref{fig:fcc2}(b) shows, its ground state is formed 
by a complete triangle with 6 monomers, a hexagon in the intermediate layer with 7 monomers, and an
incomplete triangle (possessing a ``hole'' at a corner) with 5 monomers (``6--7--5'' structure). 
Although this 
imperfection seems to destroy all rotational symmetries, it is compensated by an additional symmetry: 
Exchanging any of the triangle corners with the hole does not change the conformation at all!
Thus, the seeming imperfection has a similar effect as the energetic excitation and causes a trivial 
entropic transition. This explains, at least partly, why the 18-mer exclusively exhibits an additional 
peak in the specific heat at very low temperatures [see Fig.~\ref{fig:fcc1}(a)]. A similar
reasoning presumably also applies to the anomalous low-temperature peaks of
the 32-, 46-, and 56-mers, but for these larger ground-state conformations it does
not make much sense to go into such intricate details. 

The expectation that the 19-mer, which can form a perfect shape without any ``holes'' 
(``6--7--6'' structure), 
is a prototype 
of peculiar behavior is, however, wrong. This is due to the existence of degenerate less symmetric
ground-state conformations 
[as the exemplified conformation in Fig.~\ref{fig:fcc2}(c)].

The described geometric peculiarities are, however, only properties of very short chains.
One of the largest of the ``small'' chains that still possesses a non-spherical ground state,
is the 27-mer with the ground-state conformation  shown in Fig.~\ref{fig:fcc2}(d). 
For larger systems, the relative importance of the interior monomers will increase, because of the
larger number of possible contacts. This requires the number of surface monomers to be as small as
possible which results in compact, sphere-like shapes. A representative example is the 30-mer shown in 
Fig.~\ref{fig:fcc2}(e).
\subsection{\label{sec:theta}The $\Theta$ transition revisited}
The scaling behavior of several quantities at and close to the $\Theta$
point in three dimensions has been the subject of a large number of field-theoretic and computational
studies~\cite{grass1,grass2,grass3,pana1,yan1,mccrack1,bruns1,kremer1,mei1,kremer2,taylor1,prell3}. 
Nonetheless, the somewhat annoying result is that the nature
of this phase transition is not yet completely understood. The associated tricritical
$\lim_{n\to 0}O(n)$ field theory has an upper critical dimension $d_c=3$,
but the predicted logarithmic corrections~\cite{duplantier1,duplantier2,stephen1} could not yet
be clearly confirmed from the numerical data produced so far. In our study
of freezing and collapse on regular lattices, 
we mainly focused on the critical temperature $T_\Theta$ for polymers on the 
sc and on the fcc lattice. The sc value of $T_\Theta$ has already been precisely estimated in
several studies, but only a few values are known for the fcc case. Some previous
estimates in the literature are compiled in Table~\ref{tab:1}.
\begin{table}[t]
\caption{\label{tab:1}$T_\Theta$ values on the sc and fcc lattice from literature.}
\begin{ruledtabular}
\begin{tabular}{l@{\hspace*{4mm}}rcl@{\hspace*{4mm}}l@{\hspace*{4mm}}l}
lattice type & & $T_\Theta$ & & model & Ref.\\ 
\hline
sc  & $3.64$ & $\ldots$ & $4.13$& single chain & \cite{mccrack1}\\
    & $3.713$ & $\pm$ & $0.007$ & single chain & \cite{bruns1}\\
    & $3.650$ & $\pm$ & $0.08$  & single chain & \cite{mei1}\\
    & $3.716$ & $\pm$ & $0.007$ & single chain & \cite{grass1}\\
    & $3.60$ & $\pm$ & $0.05$\footnote{Originally given as $\beta_\Theta=0.2779\pm 0.0041$~\cite{whitt1}.}
 & single chain & \cite{whitt1}\\
    & $3.62$ & $\pm$ & $0.08$\footnote{In Ref.~\cite{whitt2} given as $\beta_\Theta=0.276\pm 0.006$.} 
& single chain & \cite{whitt2}\\
    & $3.717$ & $\pm$ & $0.003$ & single chain & \cite{grass2}\\
    & $3.717$ & $\pm$ & $0.002$ & polymer solution &\cite{grass3}\\
    & $3.745$ &       &         & lattice theory &\cite{taylor1}\\
    & $3.71$ & $\pm$ & $0.01$   & polymer solution & \cite{pana1,yan1}\\ \hline
fcc & $8.06$ & $\ldots$ & $9.43$& single chain & \cite{mccrack1}\\ 
    & $8.20$ & $\pm$ & $0.02$   & single chain & \cite{kremer1,kremer2}\\
    & $8.264$&       &          & lattice theory & \cite{taylor1}\\
\end{tabular}
\end{ruledtabular}
\end{table}

As our main interest is devoted to the expected difference of the collapse and freezing temperatures,
we will focus here on the scaling behavior of the finite-size deviation of the maximum 
specific-heat temperature of a finite-length polymer from the $\Theta$ temperature, $T_c(N)-T_\Theta$,
as it has also been studied for the bond-fluctuation model~\cite{binder1,binder1b} and the off-lattice
FENE polymer~\cite{parsons1}, as well as for polymer solution models~\cite{binder2,pana1,yan1}.
In the latter case, Flory--Huggins mean-field theory~\cite{flory1} suggests
\begin{equation}
\label{eg:fh}
\frac{1}{T_{\rm crit}(N)}-\frac{1}{T_\Theta}\sim \frac{1}{\sqrt{N}}+\frac{1}{2N},
\end{equation}
where $T_{\rm crit}(N)$ is the critical temperature of a solution of chains of 
finite length $N$ and $T_\Theta=\lim_{N\to\infty}T_{\rm crit}(N)$ is the collapse
transition temperature. In this case, field theory~\cite{duplantier1} predicts a multiplicative
logarithmic correction
of the form $T_{\rm crit}(N)-T_\Theta\sim N^{-1/2}[\ln\, N]^{-3/11}$. 
Logarithmic corrections to the mean-field theory of single chains are known, for example,
for the finite-chain Boyle 
temperature $T_B(N)$, where the second virial coefficient vanishes. The scaling of the 
deviation of $T_B(N)$ from $T_\Theta$ reads~\cite{grass1}:
\begin{equation}
\label{eq:boyle}
T_B(N)-T_\Theta\sim \frac{1}{\sqrt{N}(\ln\,N)^{7/11}}.
\end{equation}
In Ref.~\cite{parsons1}, it is claimed that, for their data obtained from simulations with the FENE potential,
this expression can also be used as a fit ansatz for
$T_c(N)-T_\Theta$.
However, also the mean-field-motivated
fit without explicit logarithmic corrections, 
\begin{equation}
\label{eq:mffit}
T_c(N)-T_\Theta=\frac{a_1}{\sqrt{N}}+\frac{a_2}{N},
\end{equation}
has been found to be consistent with the off-lattice data~\cite{parsons1}, and also with the 
results obtained by means of the bond-fluctuation model of single chains with up to 512 
monomers~\cite{binder1,binder1b}.
Up to corrections of order $N^{-3/2}$, Eq.~(\ref{eq:mffit}) is equivalent to
\begin{equation}
\label{eq:fhfit}
\frac{1}{T_c(N)}-\frac{1}{T_\Theta}=\frac{\tilde{a}_1}{\sqrt{N}}+\frac{\tilde{a}_2}{N},
\end{equation}
which was found to be consistent with numerical data obtained in grandcanonical analyses of lattice 
homopolymers and the bond-fluctuation model~\cite{binder2,pana1,yan1}.
\begin{figure}
\centerline{\epsfxsize = 8.8cm \epsfbox{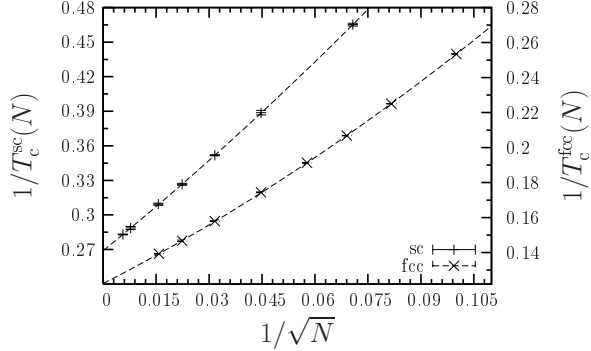}}
\caption{\label{fig:15}
Inverse collapse temperatures for several chain lengths on sc ($N\le 32\,000$) and fcc lattices
($N\le 4\,000$). Drawn lines are fits according to Eq.~(\ref{eq:fhfit}).}
\end{figure}

The situation remains diffuse as there is still no striking evidence for the predicted logarithmic 
corrections (i.e., for the field-theoretical tricritical interpretation of the $\Theta$ point) 
from experimental or numerical
data. Using our data from independent long-chain nPERMss~\cite{hsu1} chain-growth simulations 
(sc: $N_{\rm max}=32\,000$, fcc: $N_{\rm max}=4\,000$) in the vicinity of the collapse transition, 
we have performed a scaling analysis of the $N$-dependent collapse transition temperatures $T_c(N)$,
identified as the collapse peak temperatures of the individual specific-heat curves, and estimated
from it the $N\to\infty$ limit $T_\Theta$. For the single-chain system, field theory~\cite{duplantier2}
predicts the specific heat to scale at the $\Theta$ point like $C_V(T=T_\Theta)/N\sim (\ln\,N)^{3/11}$. 
Short-chain simulations~\cite{mei1} did not reveal a logarithmic behavior at all, whereas 
for long chains a scaling closer to $\ln\, N$ was read off~\cite{grass2}. 
The situation is similar for structural quantities such as the end-to-end distance and the gyration
radius. Figure~\ref{fig:15}
shows our data points of the inverse collapse temperature $T_c^{-1}$ from the simulations on the 
sc (left scale) and on the fcc lattice (right scale),
plotted against $N^{-1/2}$. Error bars for the 
individual data points in Fig.~\ref{fig:15} were obtained by jackknife error estimation~\cite{jack1} from 
several independent simulation runs. Also shown are respective fits according to the ansatz~(\ref{eq:fhfit}).
Optimal fit parameters using the data in the intervals $200\le N \le 32\,000$ (sc) and $100\le N\le 4\,000$ (fcc)
were found to be $T_\Theta^{\rm sc}=3.72(1)$, $\tilde{a}_1\approx 2.5$, and 
$\tilde{a}_2\approx 8.0$ (sc) and $T_\Theta^{\rm fcc}=8.18(2)$, 
$\tilde{a}_1\approx 1.0$, and 
$\tilde{a}_2\approx 5.5$ (fcc). In addition, we investigated also other fit functions 
motivated by field theory and
mean-field-like approaches, corresponding to Eqs.~(\ref{eg:fh})--(\ref{eq:fhfit}), each of which also with
different fit ranges. These results are listed in Tables~\ref{tab:sc} and~\ref{tab:fcc}, respectively. 
In order to decide which of the fits is consistent with our data, the $\chi^2$ test is used. Depending on
the sizes of the data sets entering into the analyses and the number of fit parameters, there are 2 to 6 
degrees of freedom $d_{\rm f}$. We make the typical assumption that deviations of the fit from the used data 
set are significant, if $\chi^2>\chi^2_{d_{\rm f};0.05}$, i.e., if $\chi^2$ lies in the $5\%$ tail of the 
$p_{d_{\rm f}}(\chi^2)$ distribution of $\chi^2$ values. In this case, with $95\%$ probability the deviations
between data and fit function are not random. The thresholds for the different degrees of freedom lie between
$\chi^2_{2;0.05}/d_{\rm f}=3.0$ and $\chi^2_{6;0.05}/d_{\rm f}=2.1$. The calculated $\chi^2$ values associated with the data 
sets and the fit functions used are also listed in Tables~\ref{tab:sc} and~\ref{tab:fcc}.

\begin{table*}
\caption{\label{tab:sc}Values of $T_\Theta$ on the sc lattice from different fits and their $\chi^2$ tests
with several degrees of freedom $d_{\rm f}$.}
\begin{ruledtabular}
\begin{tabular}{lccrcc}
fit function &  $T_\Theta$ & $a$ resp.\ $a_1$, $a_2$ & $N\hspace*{8mm}$ &$\chi^2/d_{\rm f}$ & $d_{\rm f}$\\ \hline
\multirow{3}{*}{$\displaystyle T_\text{c}(N)-T_\Theta=\frac{a}{\sqrt{N}}$}
   &  $3.6671\pm0.0051$ & $-26.0$ & 500 -- 32\,000 & 12.7 & 4\\
   &  $3.6741\pm0.0053$ & $-26.5$ & 1\,000 -- 32\,000 & 6.61 & 3\\
   &  $3.6898\pm0.0065$ & $-28.3$ & 2\,000 -- 32\,000 & 1.39 & 2\\ \hline
\multirow{3}{*}{$\displaystyle T_\text{c}(N)-T_\Theta=\frac{a}{\sqrt{N}}\,(\ln N)^{7/11}$}
   &  $3.7353\pm0.0056$ & $-8.24$ & 500 -- 32\,000 & 0.57 & 4\\
   &  $3.7370\pm0.0057$ & $-8.27$ & 1\,000 -- 32\,000 & 0.21 & 3\\
   &  $3.7398\pm0.0072$ & $-8.36$ & 2\,000 -- 32\,000 & 0.10 & 2\\ \hline
\multirow{3}{*}{$\displaystyle T_\text{c}(N)-T_\Theta=\frac{a}{\sqrt{N}\,(\ln N)^{7/11}}$}
   &  $3.6164\pm0.0048$ & $-83.0$ & 500 -- 32\,000 & 39.5 & 4\\
   &  $3.6287\pm0.0049$ & $-86.0$ & 1\,000 -- 32\,000 & 20.7 & 3\\
   &  $3.6531\pm0.0059$ & $-97.6$ & 2\,000 -- 32\,000 & 4.14 & 2\\ \hline
\multirow{3}{*}{$\displaystyle \frac{1}{T_\text{c}(N)}-\frac{1}{T_\Theta}=a\left(\frac{1}{\sqrt{N}}+\frac{1}{2N}\right)$}
   &  $3.7255\pm0.0060$ & $2.6$ & 500 -- 32\,000 & 0.28 & 4\\
   &  $3.7245\pm0.0061$ & $2.6$ & 1\,000 -- 32\,000 & 0.11 & 3\\
   &  $3.7221\pm0.0076$ & $2.6$ & 2\,000 -- 32\,000 & 0.03 & 2\\ \hline
\multirow{3}{*}{$\displaystyle \frac{1}{T_\text{c}(N)}-\frac{1}{T_\Theta}=\frac{a_1}{\sqrt{N}}+\frac{a_2}{2N}$}
   &  $3.7173\pm0.0071$ & $2.5$, $8.0$ & 200 -- 32\,000 & 0.05 & 4\\
   &  $3.7173\pm0.0104$ & $2.5$, $8.0$ & 500 -- 32\,000 & 0.07 & 3\\
   &  $3.7194\pm0.0131$ & $2.5$, $6.3$ & 1\,000 -- 32\,000 & 0.07 & 2\\ \hline
\multirow{3}{*}{$\displaystyle T_\text{c}(N)-T_\Theta=\frac{a_1}{\sqrt{N}}+\frac{a_2}{N}$}
   &  $3.7030\pm0.0059$ & $-32$, $135$ & 200 -- 32\,000 & 0.53 & 4\\
   &  $3.7090\pm0.0078$ & $-32$, $161$ & 500 -- 32\,000 & 0.25 & 3\\
   &  $3.7140\pm0.0104$ & $-33$, $186$ & 1\,000 -- 32\,000 & 0.12 & 2\\ 
\end{tabular}
\end{ruledtabular}
\end{table*}
\begin{table*}
\caption{\label{tab:fcc}Values of $T_\Theta$ on the fcc lattice using the same methodology
as in Table~\ref{tab:sc} for the sc lattice.}
\begin{ruledtabular}
\begin{tabular}{lccccc}
fit function &  $T_\Theta$ & $a$ resp.\ $a_1$, $a_2$ & $N$& $\chi^2/d_{\rm f}$ & $d_{\rm f}$\\
\hline
\multirow{5}{*}{$\displaystyle T_\text{c}(N)-T_\Theta=\frac{a}{\sqrt{N}}$}
   &  $7.2673\pm0.0052$ & $-34.1$ & 100 -- 4\,000 & 1\,000 & 6\\
   &  $7.5760\pm0.0070$ & $-39.4$ & 150 -- 4\,000 &  418 & 5\\
   &  $7.7101\pm0.0080$ & $-42.0$ & 210 -- 4\,000 &  213 & 4\\
   &  $7.8445\pm0.0096$ & $-45.5$ & 300 -- 4\,000 & 67.8 & 3\\
   &  $7.9561\pm0.0013$ & $-50.0$ & 500 -- 4\,000 & 13.6 & 2\\ \hline
\multirow{5}{*}{$\displaystyle T_\text{c}(N)-T_\Theta=\frac{a}{\sqrt{N}}\,(\ln N)^{7/11}$}
   &  $7.9218\pm0.0064$ & $-15.2$ & 100 -- 4\,000 & 200  & 6\\
   &  $8.0757\pm0.0083$ & $-16.1$ & 150 -- 4\,000 & 69.2 & 5\\
   &  $8.1356\pm0.0093$ & $-16.5$ & 210 -- 4\,000 & 32.7 & 4\\
   &  $8.1953\pm0.0110$ & $-17.0$ & 300 -- 4\,000 & 9.45 & 3\\
   &  $8.2468\pm0.0149$ & $-17.6$ & 500 -- 4\,000 & 1.00 & 2\\ \hline
\multirow{5}{*}{$\displaystyle T_\text{c}(N)-T_\Theta=\frac{a}{\sqrt{N}\,(\ln N)^{7/11}}$}
   &  $6.8260\pm0.0045$ & $-79.5$  & 100 -- 4\,000 & 2\,000 & 6\\
   &  $7.2258\pm0.0062$ & $-99.3$  & 150 -- 4\,000 & 1\,000 & 5\\
   &  $7.4166\pm0.0072$ & $-110.2$ & 210 -- 4\,000 &  500 & 4\\
   &  $7.6051\pm0.0087$ & $-125.7$ & 300 -- 4\,000 &  164 & 3\\
   &  $7.7544\pm0.0011$ & $-146.3$ & 500 -- 4\,000 & 38.1 & 2\\ \hline
\multirow{5}{*}{$\displaystyle \frac{1}{T_\text{c}(N)}-\frac{1}{T_\Theta}=a\left(\frac{1}{\sqrt{N}}+\frac{1}{2N}\right)$}
   &  $8.5434\pm0.0110$ & $1.27$ & 100 -- 4\,000 &  111 & 6\\
   &  $8.4208\pm0.0120$ & $1.23$ & 150 -- 4\,000 & 29.6 & 5\\
   &  $8.3821\pm0.0125$ & $1.22$ & 210 -- 4\,000 & 13.5 & 4\\
   &  $8.3369\pm0.0141$ & $1.20$ & 300 -- 4\,000 & 3.00 & 3\\
   &  $8.3048\pm0.0187$ & $1.18$ & 500 -- 4\,000 & 1.15 & 2\\ \hline
\multirow{4}{*}{$\displaystyle \frac{1}{T_\text{c}(N)}-\frac{1}{T_\Theta}=\frac{a_1}{\sqrt{N}}+\frac{a_2}{2N}$}
   &  $8.1778\pm0.0169$ & $1.04$, $5.49$ & 100 -- 4\,000 & 0.81 & 5\\
   &  $8.1987\pm0.0211$ & $1.06$, $5.04$ & 150 -- 4\,000 & 0.32 & 4\\
   &  $8.2107\pm0.0259$ & $1.07$, $4.75$ & 210 -- 4\,000 & 0.21 & 3\\
   &  $8.2288\pm0.0386$ & $1.09$, $4.18$ & 300 -- 4\,000 & 0.11 & 2\\ \hline
\multirow{4}{*}{$\displaystyle T_\text{c}(N)-T_\Theta=\frac{a_1}{\sqrt{N}}+\frac{a_2}{N}$}
   &  $8.0374\pm0.0110$ & $-58.9$, $360$ & 100 -- 4\,000 & 13.1 & 5\\
   &  $8.0876\pm0.0133$ & $-61.5$, $414$ & 150 -- 4\,000 & 4.71 & 4\\
   &  $8.1219\pm0.0163$ & $-63.5$, $461$ & 210 -- 4\,000 & 1.81 & 3\\
   &  $8.1640\pm0.0244$ & $-66.5$, $541$ & 300 -- 4\,000 & 0.04 & 2\\
\end{tabular}
\end{ruledtabular}
\end{table*}
From the results in Table~\ref{tab:sc} for the polymers on the sc lattice, 
we find that the two-parameter
mean-field-like fits~(\ref{eq:mffit}) and~(\ref{eq:fhfit}) as well as the single-parameter fit according 
to~(\ref{eg:fh}) are consistent with our data. Surprisingly poor, on the other hand, is the goodness of the
fit against the logarithmic scaling~(\ref{eq:boyle}). Even more astonishing is, however, the good coincidence
with a logarithmic fit of the ``wrong'' form $N^{-1/2}(\ln\, N)^{7/11}$ with the data. Summarizing these results, 
if logarithmic corrections as
predicted by tricritical field theory are present at all, even chain lengths $N=32\,000$ on an sc lattice 
are too small to observe deviations from the mean-field picture. At least, the goodness of the logarithmic
fit with the ``wrong'' exponent $+7/11$ could lead to the speculative conclusion that for 
$N\le 32\,000$ multiplicative and 
additive logarithmic corrections to 
scaling are hidden in the fit parameters of the ``mean-field-like fits''. The subleading additive
corrections are expected to be of the form $\ln(\ln\,N)/\ln^2N$~\cite{hager1}. They thus not 
only disappear very
slowly -- they are also even of the same size as the leading scaling behavior, which makes it extremely 
unlikely to observe the logarithmic corrections in computational studies at all~\cite{hager1}. 
Similar additive logarithmic scaling is also known, for example, from studies of the two-dimensional XY 
spin model~\cite{wj97}. 
The estimated sc $\Theta$ temperatures
from the good fits are in perfect agreement with the most reliable estimates from literature.

The corresponding fcc results are listed in Table~\ref{tab:fcc}. In this case, only the fit 
function~(\ref{eq:fhfit}) is independent of the data sets used and, therefore, 
consistent with the data obtained
for all chain lengths. However, the noticeable improvement of the goodness for the fits 
to~(\ref{eq:mffit}), (\ref{eg:fh}) and the ``wrong'' $N^{-1/2}(\ln\,N)^{7/11}$ form by excluding 
the very short chains from the
data sets considered, leads to the conclusion that even chains with $N=4\,000$ monomers on the fcc lattice
are also too short to find evidence for the logarithmic corrections to mean-field scaling. Our best estimates
for the fcc $\Theta$ temperature agree nicely with the results from Refs.~\cite{kremer1,kremer2}.
\section{\label{sec:summary}Summary}
Employing sophisticated chain-growth algorithms, we have performed computer simulations of 
homopolymers on sc and fcc lattices in order to analyze freezing and collapse of these chains.
Particular attention has been devoted to the question whether these transitions fall
together in the thermodynamic limit as it was reported recently from similar studies
of a specific bond-fluctuation model. In our analysis, we focus on the shifts of the
specific-heat peaks in dependence of the chain-lengths considered.

For polymers on the sc lattice, we find a remarkably systematic pattern of the 
freezing transition which can be explained
by lattice effects of the finite-length systems. In fact, the high precision of our data 
allows us to reveal a noticeable difference in the behavior of ``magic'' chain lengths that allow for
cubic or cuboid conformations. In these cases, an energy gap exists between the ground-state
conformations and the first excitations. This peculiarity causes a first-order-like 
pseudotransition which is typically more pronounced than the separate freezing transition.
Surprisingly, this effect vanishes widely for polymers with slightly longer chain lengths.
The freezing temperature decreases with increasing chain length until the next ``magic''
length is reached. Polymers on the fcc lattice behave similarly, but the relevant geometries 
are more complex. 

We have also performed comprising analyses of the collapse temperature deviations for
finite-length polymers from the $\Theta$ temperature, i.e., the collapse transition temperature
in the limit of infinite chain-length. We studied chains with lengths
of up to 32\,000 (sc) and 4\,000 (fcc) monomers, respectively. The thus obtained data were fitted
against several fit functions motivated by field-theoretic and mean-field-like approaches. 
For the chain lengths studied, we find no evidence for logarithmic
corrections as predicted by tricritical field theory. We conclude that the chain lengths are
still too short to uniquely identify logarithmic corrections which are probably effectively
taken into account by the
amplitudes of the dominant mean-field terms. 

From our results for the freezing and the collapse transition, we conclude that both transitions
remain well separated also in the extrapolation towards the thermodynamic limit. This is 
the expected behavior as it is a consequence of the extremely short range of attraction in the 
nearest-neighbor lattice models 
used. 
Considering a more general square-well contact potential between nonbonded monomers
in our parametrization,
\begin{equation}
v(r)=\left\{\begin{array}{rl}
\infty & \quad r\le 1,\\
-1 &  \quad 1<r\le \lambda,\\
0 & \quad \lambda<r,
\end{array}\right.
\end{equation}
the attractive interaction range is simply 
$R=\lambda-1$. 
In our single-chain study of sc and fcc lattice models, we have
$\lambda\to 1$ and thus $R\to 0$. Since this $R$ value is well below a crossover threshold
known for colloids interacting via Lennard--Jones-like and Yukawa potentials, where 
different solid phases can coexist, $R_c^{(1)}\approx 0.01$~\cite{baus1,frenkel2,baus2}, we
interpret our low-temperature transition as the restructuring or ``freezing'' of 
compact globular shapes into the (widely amorphous) polymer 
crystals.

Following Ref.~\cite{frenkel1}, there is also \emph{another} phase boundary, namely
between stable and metastable colloidal
vapor-liquid (or coil-globule) transitions, in the range $0.13 < R_c^{(2)} < 0.15$. 
Other theoretical and experimental approaches yield slightly larger values, 
$R_c^{(1)}\approx 0.25$~\cite{baus1,rascon1,ilett1,asherie1}.
Below $R_c^{(2)}$, the liquid (globule) phase is only metastable. The
specific bond-fluctuation model used in Ref.~\cite{binder1} corresponds to $R=0.225$, i.e., it lies in the 
crossover regime between the stable and metastable liquid phase~\cite{binder1b}. 
Consequently, the crystallization and collapse transition merge in the 
infinite-chain limit and
a stable liquid phase was only found in a subsequent study of a 
bond-fluctuation model with larger interaction range~\cite{binder1b}.

Qualitatively, analogous to the behavior of colloids, our considerations
would explain the separate stable crystal, globule, and random-coil
\mbox{(pseudo)phases} that we have clearly identified in our lattice polymer study. 
Since the range of interactions seems to play a crucial, quantitative role,
it is 
an interesting, still widely open question to what extent the colloidal picture 
in the compact crystalline and globular phases is systematically
modified for polymers with different nonbonded interaction ranges, where steric constraints 
(through covalent bonds) are a priori not negligible. 
\section{Acknowledgments}
We thank K.\ Binder, W.\ Paul, F.\ Rampf, and T.\ Strauch for helpful discussions.
This work is partially supported by the German Science Foundation (DFG) under
Grant No.\ JA 483/24-1/2 and by a computer grant of the John von Neumann Institute for 
Computing (NIC), Forschungszentrum
J\"ulich, under No.\ hlz11. M.B.\ acknowledges support by research fellowships of the
DFG under Grant No.\ BA 2878/1-1 and the Wenner-Gren Foundation. We are also grateful for support by 
the DAAD-STINT personnel exchange program between Germany and Sweden.

\begin{thebibliography}{99}
\bibitem{rem1}
This gap is an artifact of the simple-cubic lattice and the resulting excitation transition
is a non-cooperative effect: Consider a cuboid conformation with a chain end located in one of
the corners. It forms two energetic contacts with the nearest neighbors nonadjacent 
within the chain. Performing a local off-cube pivot-rotation of the bond the chain end is 
connected with, these two contacts are lost and none new is formed.
\end{thebibliography}
\end{document}